\documentclass[aps,superscriptaddress,twocolumn,twoside,floatfix,pra,a4paper]{revtex4-1} 
\usepackage{times}
\usepackage{epsfig}
\usepackage{amsfonts}
\usepackage{amsmath}
\usepackage{amssymb,amsthm}
\usepackage{color}
\usepackage{multirow}
\usepackage{braket}
\usepackage{latexsym}
\usepackage{amsfonts}
\usepackage{mathrsfs}
\usepackage{natbib}
\usepackage{verbatim}
\usepackage{gensymb}

\usepackage[colorlinks=true,linkcolor=blue,citecolor=magenta,urlcolor=blue]{hyperref}

\theoremstyle{plain}

\def\bea{\begin{eqnarray}}
\def\eea{\end{eqnarray}}
\def\ba{\begin{array}}
\def\ea{\end{array}}

\def\beq{\begin{equation}}
\def\eeq{\end{equation}}

\newcommand{\ketbra}[2]{|#1\rangle \langle #2|}

\begin{document}

\title{Self-testing of binary Pauli measurements requiring neither entanglement nor any dimensional restriction}

\author{Ananda G. Maity}
\email{anandamaity289@gmail.com}
\affiliation{S. N. Bose National Centre for Basic Sciences, Block JD, Sector III, Salt Lake, Kolkata 700 106, India}

\author{Shiladitya Mal}
\email{shiladityamal@hri.res.in}
\affiliation{Harish-Chandra Research Institute, HBNI, Chhatnag Road, Jhunsi, Allahabad 211 019, India}
\affiliation{Department of Physics and Center for Quantum Frontiers of Research and Technology (QFort), National Cheng Kung University, Tainan 701, Taiwan}

\author{Chellasamy Jebarathinam}
\email{jebarathinam@cft.edu.pl}
\affiliation{Department of Physics and Center for Quantum Frontiers of Research and Technology (QFort), National Cheng Kung University, Tainan 701, Taiwan}
\affiliation{Center for Theoretical Physics, Polish Academy of Sciences, Aleja Lotnik\'ow 32/46, 02-668 Warsaw, Poland}

\author{A. S. Majumdar}
\email{archan@bose.res.in}
\affiliation{S. N. Bose National Centre for Basic Sciences, Block JD, Sector III, Salt Lake, Kolkata 700 106, India}

\begin{abstract}
Characterization of quantum devices received from unknown providers is a significant primary task for any quantum information processing protocol. Self-testing protocols are designed for this purpose of certifying quantum components from the observed statistics under a set of minimal assumptions. Here we propose a self-testing protocol for certifying binary Pauli measurements employing the violation of a Leggett-Garg inequality. The scenario based on temporal correlations does not require entanglement, a costly and fragile resource. Moreover, unlike  previously proposed self-testing protocols in the prepare and measure scenario, our approach requires neither dimensional restrictions, nor other stringent assumptions on the type of measurements. We further analyse the robustness of this hitherto unexplored domain of self-testing of measurements.
\end{abstract}

\maketitle

\section{Introduction}
It is hard to overemphasise the role of emerging quantum technology in recent times. Various real life applications, such as quantum key distribution \cite{qkd}, quantum sensing \cite{qsensing}, quantum metrology \cite{metrology}, quantum internet \cite{internet}, and machine learning \cite{mlearning} have been  investigated with great prospects. For such prospects to materialize in practice, it becomes utmost important to design experiments which can test whether the quantum components of the required devices function properly. To this end, various certification or verification protocols have been employed, such as those based on tomography and benchmarking \cite{bench}, as well as self-testing protocols based essentially on observed statistics \cite{my'98, my'04}. 

Among several certification protocols generic quantum tomography is the most powerful, but at the same time it is the most resource consuming. Randomized benchmarking refers to a collection of methods that aim at reliably estimating the figure of merit of the overlap between the physical quantum process and it's ideal counterpart. For both randomized benchmarking or quantum tomography, trusted mesurements are required in order to certify quantum devices. On the other hand, self-testing of quantum components based on  statistics collected from lesser number of measurements and minimal physical assumptions, require lesser trust on the measurement devices and are experimentally more efficient and less resource consuming\cite{bench}.

Self-testing based on the Bell test was proposed initially in the context of quantum cryptography \cite{my'98, my'04},  where it was shown that maximum violation of the Bell-CHSH inequality \cite{Bell'64,chsh'69} implies that the underlying quantum state is maximally entangled and the local measurements performed by two distant parties are anticommuting \cite{popescu'92}. In such self-testing protocols one can uniquely identify quantum states and measurements (upto some local isometries) by observing extreme correlations in measurement statistics. The underlying assumptions are the same as those required for implementation of loophole free Bell tests \cite{hensen'15,knill'15,zeilinger'15,weinfurter'17}. Since then all pure bipartite entangled states and certain multipartite quantum states such as graph states have been self-tested \cite{Col,mayers2003} in this scenario. Self-testing of pure entangled bipartite states has also been investigated \cite{Supic,goswami,Goh,SBK20} employing Einstein-Podolski-Rosen steering \cite{Wiseman}. Self-testing using steering is weaker in the sense that it requires more assumptions compared to that based on Bell tests, but the former is easier to verify experimentally \cite{peng}. 
 
In this regard, it is worthwhile to mention that self-testing based on maximal violation of some nonclassicality revealing criterion is very fragile as  idealised situations in the laboratory are rare to occur. Therefore, a self-testing statement is useful for practical purposes if it has reasonable robustness \cite{Scarani2,Yang,Bamps, Kani2}. For instance, there are robust self-testing protocols for which given a certain level of violation of a Bell inequality (but not necessarily maximal), nontrivial lower bound on the fidelity between the initially unknown state and a given target state has been shown. Robust self-testing of higher dimensional state has also been studied \cite{Scarani3,supic2018,Zhang'18}. Apart from various special class of quantum states, different types of measurements including maximally incompatible observables \cite{popescu'92}, entangling Bell-measurement \cite{bancal'18,renou'18}, binary observables \cite{kani'17}, more than two dichotomic observables \cite{Mck,Bowles'18}, and non-maximally anticommuting observables \cite{Pironio} have also been self-tested (See Ref.\cite{supic'19} for a review).
 
Apart from the above schemes of self-testing of measurements based on the use of quantum correlations in entangled states, some certification methods for self-testing have been proposed that do not require entanglement \cite{chen'11,brunner'15,bourennane'16,Tavakoli}. A pair of anticommuting Pauli observables  have been self-tested in the prepare and measure scenario \cite{tavakoli'18}, employing the quantum advantage of $2\mapsto 1$ random access code (RAC) \cite{ambainis'99, tavakoli'15}. Such approaches in the context of self-testing of local states and measurements are more resource efficient and experimentally feasible compared to schemes based on nonlocal quantum correlations since they do not require entanglement, an expensive resource. However, the assumption of an upper bound on the system-dimension has been crucial for such approaches, and hence, the verifier needs to trust the measurement devices more. 

Another self-testing protocol has been constructed for three-dimensional states and measurements \cite{kishor'19} based on contextuality of quantum theory \cite{ks'67}, which requires certain restrictive assumptions such as the compatibility  between  subsequent projective measurements \cite{supic'19}. Recently, an extension of the scheme has been proposed based on the sum of sqaures decomposition of a family of non-contextual inequalities \cite{kcbs}, relaxing some of the above restrictions \cite{saha'20}. Self-testing of arbitrary high-dimensional local quantum systems using contextuality has also been proposed where the assumption of projective measurement is necessary \cite{kishor'19(1)}. The above protocols pertain to local measurements in three and above dimensions. On the other hand, among all measurements in quantum information processing, the measurements with dichotomic settings and outcomes are the most widely used ones. Hence, the design of resource efficient self-testing protocols for such meaurements with minimal assumptions is necessary for practical purposes.

In the present work, we develop a scheme of self-testing binary Pauli measurements in a hitherto unexplored scenario, {\it i.e.}, employing nonclassicality of temporal correlations exhibited via violation of the Leggett-Garg inequality(LGI) \cite{LG'85,emary'14} and predictability\cite{kofler'13,rand'16}, which does not require any entanglement. LGI has been employed to probe macroscopic coherence and to study quantum to classical transition \cite{LG'85,emary'14,LG'02, kofler'07,kofler'13,brukner'04,mal'16, mal'18,mal'19}. Various aspects of temporal correlations have been investigated \cite{fritz'10,budroni'13,rand'16,bm'14,brierley'19,usha'13,alok'17,das'18,ku'18,titas'18} and experimentally realized \cite{knee'12,robens'15,knee'16,ku'19,shayan'19,spee'20}. Here, we employ the violation of LGI,  and invoke a minimal set of assumptions which can be easily met in a real experiment, in order to provide a self-testing statement for binary Pauli measurements without any need to restrict the dimension of the Hilbert space accessed by the quantum devices. We further perform the robustness analysis of our self-testing protocol by deriving a lower bound on the fidelity of the measured observables with that of ideal ones.

\section{Description of the scenario}
\begin{figure}
\begin{center}
\includegraphics[width=0.42\textwidth]{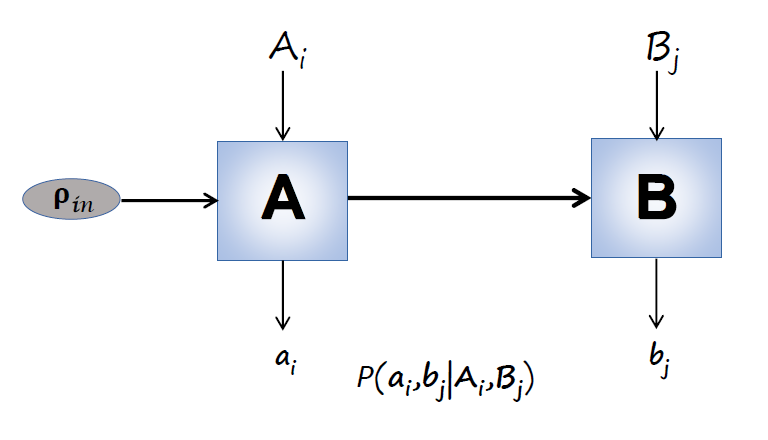}
\caption{\textcolor{black}{Sequential-measurement set-up. The joint probability distribution $P(a_i,b_j \mid A_i,B_j)$, {\it i.e.}, the probability of obtaining outcome $a_i$ and $b_j$ when Alice measures $A_i$ in an arbitrary input state at some instant $t_m$, and Bob performs measurement $B_j$ at some later instant $t_{m+1}$, respectively, is observed to obtain the violation of  LGI (\ref{k4}).} \label{Seq_LGI}}
\end{center}
\end{figure}

We start with a brief description of the Leggett-Garg (LG) test (Fig.-\ref{Seq_LGI}) which enables self-testing of Pauli measurements whenever extremal correlations of LGI are observed. In a LG test, a single system is measured sequentially at different instants of time in order to obtain temporal correlations. Two sequential binary measurements are performed (by Alice and Bob, respectively) on an identical initial state prepared by the experimenter in every run of the experiment. In contrast to the sequential measurement scenario of self-testing using contextuality \cite{kishor'19}, here subsequent measurements are not required to commute with each other.  

Alice and Bob have two choices of binary measurements, say, $\{A_1,A_2\}$ and $\{B_1,B_2\}$ to perform in each run. The probability of obtaining outcome $a_i$ and $b_j$ is denoted by $P(a_i,b_j \mid A_i,B_j)$, when Alice measures $A_i$ at time $t_m$ and Bob measures $B_j$ at some later instant $t_{m+1}$, with $i,j \in\{1,2\}$ and $a_i, b_j\in\{0,1\}$. Let us denote, $\mathcal{P}_{a_i|A_i}$, $\mathcal{P}_{b_j|B_j}$ as projectors so that $\sum_{a_i}\mathcal{P}_{a_i|A_i}=\mathbb{I}, \sum_{b_j}\mathcal{P}_{b_j|B_j}=\mathbb{I}$. The two-time joint probability can be obtained using Bayes' rule as,
\begin{align}\label{pij}
&P(a_i,b_j \mid A_i,B_j) = P(a_i \mid A_i) P(b_j\mid a_i, A_i,B_j ) \nonumber \\
&= \text{Tr}\left[ \mathcal{P}_{a_i|A_i} \rho_{in}\right] \text{Tr}\left[ \mathcal{P}_{b_j|B_j} \frac{\mathcal{P}_{a_i|A_i}\rho_{in}\mathcal{P}_{a_i|A_i}^{\dagger}}{\text{Tr}\left[ \mathcal{P}_{a_i|A_i}\rho_{in}\mathcal{P}_{a_i|A_i}^{\dagger}\right]}\right] .
\end{align}
The two-time correlation is defined as,
\begin{equation}\label{cij}
\mathcal{C}_{ij} =\sum_{a_i,b_j} (-1)^{a_i\oplus b_j} P(a_i,b_j \mid A_i,B_j),
\end{equation}
where $\oplus$ denotes addition modulo $2$. 
 The four-term LGI in terms of the above correlators is given by,
\begin{equation}\label{k4}
\mathcal{K}_4 = C_{11} + C_{21} + C_{22} - C_{12} \leq 2.
\end{equation}
Any model compatible with classical theory predicts that the maximal value of the above expression is $2$, whereas the quantum theory can violate this inequality up to $2\sqrt{2}$. Suppose, in a single run, Alice performs a projective qubit measurement, $A_1\equiv \hat{a_1}.\vec{\sigma}$ (to make the preliminary discussion simple, but it can be any general measurement too) on an arbitrary input state, $\rho_{in}$ at time $t_1$, and Bob performs a qubit measurement, $B_1\equiv \hat{b_1}.\vec{\sigma}$ at some later time $t_2$, where $\hat{a_1}$ and $\hat{b_1}$ are the Bloch vectors denoting Alice's and Bob's measurement directions respectively and $\vec{\sigma}$ is the vector of Pauli matrices. Then the maximum quantum violation, $2\sqrt{2}$, can be achieved by the following measurement settings,
\begin{eqnarray}\label{mea}
A_1^{\text{ideal}}= \sigma_z, \nonumber \\
A_2^{\text{ideal}}=  \sigma_x, \nonumber \\
B_1^{\text{ideal}} = \frac{\sigma_x + \sigma_z}{\sqrt{2}}, \nonumber \\
B_2^{\text{ideal}} =  \frac{\sigma_x - \sigma_z}{\sqrt{2}},
\end{eqnarray}
upto some local unitary. In the context of self-testing, all the four measurements are {\it a priori} unknown to Alice and Bob. Our goal is to certify above measurements from the observed statistics, under some suitable and minimal assumptions.

\section{Derivation of Leggett-Garg inequality}
LGI has been earlier derived using the assumptions of non-invasive measurement and realism \cite{brukner'04,mal'16}. However, those assumptions are at ontological level, and hence, cannot be verified individually. Rather a conjunction of them can be verified in an experiment. On the other hand, in \cite{rand'16}, LGI was derived from some operational assumptions called predictability and no signalling in time (NSIT) \cite{kofler'13}. Subsequently, the NSIT condition has been experimentally observed on a variety of input states \cite{expnsit}. Here we present the essential features of the LGI derivation \cite{rand'16}. Let us first state the two assumptions in precise mathematical form.

\textbf{Predictability :} A model is said to be predictable if the joint statistics $P(a_i,b_j \mid A_i,B_j)\in\{0,1\}~ \forall a_i,b_j,A_i,B_j$ \cite{wiseman'12}.
\\
\textbf{NSIT :} NSIT is defined by the condition that measurement statistics is not influenced by the earlier measurements. Mathematically, $P(b_j\mid B_j)=P(b_j\mid A_i, B_j) ~\forall A_i, B_j,b_j$. 
\\
In order to derive LGI from the above two assumptions, our aim is to show that
\begin{eqnarray}
NSIT \wedge Predictability \Rightarrow LGI.
\end{eqnarray}
Suppose $\lambda$ denotes some underlying variable at the ontological level, averaging over which we obtain joint probabilities observed in an experiment, {\it i.e.}, 
\begin{eqnarray}
P(a_i,b_j \mid A_i,B_j)=\int_{\lambda} d\lambda p(\lambda) P(a_i,b_j \mid A_i,B_j,\lambda). \nonumber
\end{eqnarray}
LGI follows in a straightforward way when the joint probability at the ontological level gets factorised, or in other words, we have to show that predictability together with NSIT leads to 
\begin{eqnarray}\label{facto}
P(a_i,b_j \mid A_i,B_j,\lambda)=P(a_i\mid A_i,\lambda) P(b_j \mid B_j,\lambda).
\end{eqnarray}
As further conditioning does not change the deterministic probability distribution, we have from the predictability, $P(a_i,b_j \mid A_i,B_j,\lambda)=P(a_i,b_j \mid A_i,B_j)$. Using Bayes' rule one  has 
\begin{equation}
P(a_i,b_j \mid A_i,B_j)=P(a_i\mid A_i, B_j, b_j) P(b_j\mid A_i, B_j). \nonumber
\end{equation}
The condition of NSIT implies $P(b_j\mid A_i, B_j)=P(b_j\mid B_j)$. Also, as
physically reasonable and broadly accepted, a later measurement cannot influence the past measurement result, and hence $P(a_i\mid A_i, B_j, b_j)=P(a_i\mid A_i)$. With the above implications, one can construct a theory at the ontological level where  $P(a_i\mid A_i,\lambda)=P(a_i\mid A_i)$, $P(b_j\mid B_j,\lambda)=P(b_j\mid B_j)$, which lead to Eq. \eqref{facto}. Now, a straightforward calculation leads us to derive the four-term LGI \eqref{k4} which is bounded by $2$.

Despite some structural similarities, the LG-test is essentially different from the Bell-test \cite{clemente}, and also has some loopholes which are different from that of the Bell-test \cite{wilde}. The approach adopted in this paper for the derivation of LGI is based on the conjunction of {\it Predictability} and {\it NSIT}. In the present scenario it is ensured that NSIT condition is satisfied, as we see in the next section. This means that if LGI is violated, then predictability must have to be violated. Thus, the violation of predictability naturally guarantees non-classicality which is required for the purpose of self-testing.

\section{Self-testing of measurements using Leggett-Garg inequality} 
The above derivation of LGI with the assumptions of predictability and NSIT helps us to devise a self-testing protocol for binary Pauli measurements \eqref{mea} whenever extremal non-classical temporal correlations are observed. We formulate the self-testing protocol in such a way that the NSIT condition is satisfied so that maximal violation of the LGI will imply the violation of the predictability condition, and hence, no classical strategy is able to reproduce the statistics. We now make a minimal assumption which is again very natural in the sequential measurement scenario, since otherwise, a classical model may simulate the quantum violation of LGI.

\textbf{Assumption :}~\emph{The measurement device of Alice acts only on the input state prepared by the experimenter, and the measurement device of Bob acts only on the state produced by Alice's measurement, with both returning only the respective post-measurement states.}

A similar assumption was also considered in Ref.\cite{saha'20} for the purpose of self-testing of local three dimensional measurements using the sum-of-squares decomposition of a family of non-contextual inequalities \cite{kcbs}. However, in our approach using LGI we are interested here in the self-testing of binary qubit measurements. Employing the maximum violation of LGI along with NSIT condition we can self-test the binary qubit measurements given by Eq. \eqref{mea} upto some local unitaries. Our self-testing protocol does not depend on the input state. However, for the sake of a formal proof of self-testing, we first consider a general qubit state (in Lemma 1). Then, making use of this proof, we extend our result for states in any  dimension (Lemma 2). Our formal proof of self-testing is thus accomplished in the following two steps (lemma 1 and 2), and finally we present the proof of isometry in Theorem 1. 

\textbf{Lemma 1.}~ \emph{The maximum violation of LGI (i.e, $\mathcal{K}_4^{\text{max}} = 2\sqrt{2}$) implies implementation of the qubit measurement observables given by equation \eqref{mea} upto some local unitaries, satisfying the NSIT condition. }

\begin{proof} Maximum quantum violation of LGI can only be achieved if the measurements are taken to be projective . We maximize the LGI considering the most general positive operator valued measure (POVM) with two outcomes. Maximizing the four-time LGI expression numerically over all the POVM parameters, it is found that the maximum value, $2\sqrt{2}$ can only be achieved if the measurements are taken to be projective (the details of the proof are given in  Appendix-\ref{Appa}). Hence, without any loss of generality, we restrict ourselves to projective measurements only.		
\\
To obtain the two-time correlation $\mathcal{C}_{ij}$ we have to calculate joint probabilities $P(a_i,b_j \mid A_i, B_j)$ given in Eq.\eqref{pij}. Considering the  general qubit input state, $\rho_{in}=\frac{\mathbb{I}+\hat{n}.\vec{\sigma}}{2}$, the first term of Eq.\eqref{pij} can be simplified to $\text{Tr}\left[ \mathcal{P}_{a_i|A_i} \rho_{in}\right] = \frac{1}{2}(1+(-1)^{a_i}\hat{a_i}.\hat{n})$. The post-measurement state after obtaining outcome $a_i$ is given by, $\frac{\mathcal{P}_{a_i|A_i}\rho_{in}\mathcal{P}_{a_i|A_i}^{\dagger}}{\text{Tr}\left[\mathcal{P}_{a_i|A_i}\rho_{in}\mathcal{P}_{a_i|A_i}^{\dagger} \right] } = \frac{1}{2}\left[ \mathbb{I}+ \left( -1\right) ^{a_i}\hat{a_i}.\vec{\sigma}\right]$.
Now, the term, $\text{Tr}\left[\mathcal{P}_{b_j|B_j}\frac{\mathcal{P}_{a_i|A_i}\rho_{in}\mathcal{P}_{a_i|A_i}^{\dagger}}{\text{Tr}\left[ \mathcal{P}_{a_i|A_i}\rho_{in}\mathcal{P}_{a_i|A_i}^{\dagger}\right]}\right]$ can be simplified to $\frac{1}{2}\left[ 1+ (-1)^{a_i+b_j}\hat{a_i}.\hat{b_j}\right]$.
So, the joint probability distribution of getting outcomes $a_i$ and $b_j$ when measurements $A_i$ and  $B_j$ are performed respectively, is given by,
\begin{equation}\label{jp}
P(a_i,b_j \mid A_i, B_j)= \frac{1}{4}(1+(-1)^{a_i}\hat{a_i}.\hat{n})( 1+ (-1)^{a_i+b_j}\hat{a_i}.\hat{b_j}).
\end{equation}
After simplification, we get, $\mathcal{C}_{ij}= \hat{a_i} . \hat{b_j}.$ Then, the four-time LGI in terms of four correlators will be,
\begin{eqnarray}
\mathcal{K}_4 &= \hat{a_1} . \hat{b_1}+\hat{a_2} . \hat{b_1}+\hat{a_2} . \hat{b_2} -\hat{a_1} . \hat{b_2}\leq 2.
\end{eqnarray}
It follows that the maximum value of $\mathcal{K}_{4}$ is $2\sqrt{2}$ when all $\vert \hat{a_i}.\hat{b_j}\vert=\frac{1}{\sqrt{2}}$ \cite{fritz'10}, which can be obtained for the settings described in Eq. (4) or their local unitary rotation. 
\\	
Here, one can verify that the NSIT condition is also satisfied in case of maximum LGI violation when $\vert \hat{a_i}.\hat{b_j}\vert=\frac{1}{\sqrt{2}}$. The NSIT condition on the probabilities, $P(b_j\mid A_i, B_j)=P(b_j\mid B_j)$ $\forall b_j,A_i,B_j$, implies $(-1)^{a_1+b_1}\hat{a_1}.\hat{b_1} = (-1)^{a_2+b_1}\hat{a_2}.\hat{b_1}=(-1)^{a_1+b_2}\hat{a_1}.\hat{b_2}=(-1)^{a_2+b_2}\hat{a_2}.\hat{b_2}$. Thus clearly, when maximum violation of $\mathcal{K}_{4}$ is observed along with the NSIT condition, it is obvious that the predictability condition does not hold anymore.
\end{proof}

We now show how maximal violation of LGI can also be used to self-test the Pauli measurements even if the measurement operators act on a higher dimensional Hilbert space. Let the initial state be $\rho_{in}$ in an arbitrary dimensional Hilbert space, and the measurements $A_i, B_j$ both act on this space.

\textbf{Lemma 2.}~ \emph{The maximum violation of LGI (i.e, $\mathcal{K}_4^{\text{max}} = 2\sqrt{2}$) implies implementation of the block diagonal measurement, i.e., $A_1 = \oplus_i \sigma_z^i,   A_2= \oplus_i \sigma_x^i ,  B_1 = \oplus_j (\sigma_x^j+ \sigma_z^j)/\sqrt{2},  B_2 = \oplus_j (\sigma_x^j- \sigma_z^j)/\sqrt{2}$.}
\begin{proof}
Suppose $\left\lbrace \mathcal{P}_{a_i|A_i}\right\rbrace $ and $\left\lbrace \mathcal{P}_{b_j|B_j}\right\rbrace $ are two dichotomic measurements which act on an arbitrary dimensional Hilbert space. Then, according to Jordan's lemma \cite{Kani2}, $ \mathcal{P}_{a_i|A_i}= \oplus _m  \mathcal{P}_{a_i|A_i}^{m} $ and $ \mathcal{P}_{b_j|B_j}= \oplus _n  \mathcal{P}_{b_j|B_j}^{n} $, where $\mathcal{P}_{a_i|A_i}^{m} $ and $\mathcal{P}_{b_j|B_j}^{n}$ are projectors on $\mathcal{H}_d$ with $d\leq 2$ (here, in view of our proof of lemma 1, we again stick to projective measurements without any loss of generality). Therefore, one has
\begin{align}
&P(a_i,b_j \mid A_i,B_j) = \nonumber\\
&\sum_{m,n}p_m \text{Tr}\left[ \mathcal{P}_{a_i|A_i}^{m} \rho_{m}\right] \text{Tr}\left[ \mathcal{P}_{b_j|B_j}^{n} \frac{\mathcal{P}_{a_i|A_i}^{m}\rho_{m}\mathcal{P}_{a_i|A_i}^{m \dagger}}{\text{Tr}\left[ \mathcal{P}_{a_i|A_i}^{m}\rho_{m}\mathcal{P}_{a_i|A_i}^{m \dagger}\right]}\right] \nonumber
\end{align}
where $p_m = \text{Tr}(\rho_{in}\Pi_m)$, with $\Pi_m =\sum_{a_i}\mathcal{P}^m_{a_i|A_i}$, and $\rho_m = (\Pi_m \rho_{in}\Pi_m)/p_m$, which is at most a qubit state. with $\rho=\oplus_m p_m \rho_m$, where $\rho_m = \frac{\mathbb{I}+\hat{r}.\vec{\sigma}}{2}$ acts on $\mathcal{H}_d$ with $d\leq 2$. Now, the operators of the above equation acting on $\mathcal{H}_d$ with $d=2$  implies that
\begin{eqnarray}
&&P(a_i,b_j \mid A_i, B_j) = \nonumber \\
&& \sum_{m}p_m \frac{1}{4}(1+(-1)^{a^m_i}\hat{a^m_i}.\hat{r})( 1+ (-1)^{a_i+b_j}\hat{a^m_i} \cdot \hat{b^n_j}).\nonumber
\end{eqnarray}
From the above equation  it follows that,
\begin{eqnarray}
C_{ij}= \sum_{m,n}p_m \hat{a^m_i} \cdot \hat{b^n_j}
\end{eqnarray}
The above expectation implies $C_{11} + C_{21} + C_{22} - C_{12} = 2\sqrt{2}$ if and only if $\hat{a_1}^m = \hat{z}, \hat{a_2}^m = \hat{x},\hat{b_1}^n = (\hat{z} + \hat{x})/\sqrt{2},\hat{b_2}^n = (\hat{z} -\hat{x})/\sqrt{2}$ and $\sum_{m}p_m = 1$. This is achieved if and only if $A_1 = \oplus_m \sigma_z^m,   A_2= \oplus_m \sigma_x^m ,  B_1 = \oplus_n (\sigma_x^n+ \sigma_z^n)/\sqrt{2},  B_2 = \oplus_n (\sigma_x^n - \sigma_z^n)/\sqrt{2}$ and $\rho_{in} = \oplus_m  \rho_m$.
\end{proof}
The above two lemmas enables us to present the following theorem.

\textbf{Theorem 1.}~ \emph{If $\mathcal{K}_4^{\text{max}} = 2\sqrt{2}$ is observed in LG-test under Assumption 1, with the measurements of Alice, $A_i$ acting on $H_d $, producing the post measurement states $\left\lbrace  \frac{\mathcal{P}_{a_i|A_i}\rho_{in}\mathcal{P}_{a_i|A_i}^{\dagger}}{\text{Tr}\left[ \mathcal{P}_{a_i|A_i}\rho_{in}\mathcal{P}_{a_i|A_i}^{\dagger}\right]} \right\rbrace_a$,  and the measurements of Bob $B_j$ acting on these post measurement states, then there exists an isometry $\Phi : \mathcal{H}^d \rightarrow  \mathcal{C}^2 \otimes \mathcal{H}^d $  such that
\begin{eqnarray}
&&\Phi\left( B_j \frac{\mathcal{P}_{a_i|A_i}\rho_{in}\mathcal{P}_{a_i|A_i}^{\dagger}}{\text{Tr}\left[ \mathcal{P}_{a_i|A_i}\rho_{in}\mathcal{P}_{a_i|A_i}^{\dagger}\right]} \right) \Phi^{\dagger} \nonumber \\
&=&B^{\texttt{ideal}}_j \left|\psi^{\texttt{ideal}}_{a|A_i}\right \rangle \left \langle\psi^{\texttt{ideal}}_{a|A_i} \right |\otimes \left| \texttt{junk} \right \rangle \left \langle \texttt{junk}\right |\nonumber
\end{eqnarray}
where $\left|\psi^{\texttt{ideal}}_{a|A_i}\right \rangle$ are the eigenstates of Alice's ideal measurements and $B^{\texttt{ideal}}_j$ are Bob's ideal measurements given by Eq. (4) respectively, and $\left| \texttt{junk} \right \rangle $ is a junk state acting on $\mathcal{H}^d$.}
\begin{proof}
The details proof are given in the Appendix-\ref{Appb}. Choosing the eigenbasis of $A_1$ as the computational basis, from lemma $1$ and $2$, it follows that the post measurement states of Alice can be written as  
$\frac{\mathcal{P}_{a_i|A_i}\rho_{in}\mathcal{P}_{a_i|A_i}^{\dagger}}{\text{Tr}\left[ \mathcal{P}_{a_i|A_i}\rho_{in}\mathcal{P}_{a_i|A_i}^{\dagger}\right]} = \oplus_m p_m \left|\psi^m_{a|A_i}\right \rangle \left \langle \psi^m_{a|A_1}\right |$, with $\left|\psi^m_{0|A_1}\right \rangle= \left| 2m \right \rangle $
and $\left|\psi^m_{1|A_1}\right \rangle= \left| 2m +1 \right \rangle $. We append an ancilla qubit prepared in the state $\left|0 \right \rangle$
and look for an isometry $\Phi$ such that 
\begin{eqnarray}
&&\Phi\left( B_j \frac{\mathcal{P}_{a_i|A_i}\rho_{in}\mathcal{P}_{a_i|A_i}^{\dagger}}{\text{Tr}\left[ \mathcal{P}_{a_i|A_i}\rho_{in}\mathcal{P}_{a_i|A_i}^{\dagger}\right]} \otimes \left|0 \right \rangle \left \langle 0 \right | \right) \Phi^{\dagger} \nonumber \\
&=&B^{\texttt{ideal}}_j \left|\psi^{\texttt{ideal}}_{a|A_i}\right \rangle \left \langle\psi^{\texttt{ideal}}_{a|A_i} \right |\otimes \left| \texttt{junk} \right \rangle \left \langle \texttt{junk}\right |\nonumber
\end{eqnarray}
This can be achieved for $\Phi$ defined by the map
\begin{align}
    \Phi \left|2m,0 \right \rangle &\rightarrow  \left| 2m, 0   \right \rangle  \nonumber  \\
    \Phi \left|2m+1,0 \right \rangle &\rightarrow  \left| 2m, 1  \right \rangle \nonumber
\end{align}
\end{proof}

\section{Robust self-testing of measurements}
Robustness of self-testing of measurements is quantified as how the actual observables which are to be self-tested differ from the ideal ones. Hence, characterization of the fidelities between the real measurements and the ideal measurements is warranted. Let us first perform the robustness analysis of the measurements on Alice's side i.e, $\left\lbrace \mathcal{P}_{a_i|A_i}\right\rbrace$. Given an arbitrary set of measurements $\left\lbrace \mathcal{P}_{a_i|A_i}\right\rbrace$, the average fidelity with the ideal measurements are,
\begin{equation}
S(\left\lbrace \mathcal{P}_{a_i|A_i}\right\rbrace) = \text{max}_\Lambda \sum_{i,a_i} F(\mathcal{P}_{a_i|A_i}^{\text{ideal}}, \Lambda[\mathcal{P}_{a_i|A_i}])/4.
\end{equation}  
Here $\Lambda$ is a quantum channel and the fidelities are defined as usual, $F(\rho,\sigma)=\text{Tr}(\sqrt{\sqrt{\rho}\sigma\sqrt{\rho}})$. The fidelities between the real and the ideal measurements, $F(\mathcal{P}_{a_i|A_i}^{\text{ideal}}, \Lambda[\mathcal{P}_{a_i|A_i}])$ simplify to $\text{Tr}(\Lambda \left[ \mathcal{P}_{a_i|A_i} \right] \mathcal{P}_{a_i|A_i}^{\text{ideal}})$. The lower bound on the smallest possible value of the average fidelity $S$, given a particular violation of the four-time LGI $\mathcal{K}_4$, can be found by minimizing over all sets of four measurements of $\mathcal{P}_{a_i|A_i}$,
\begin{equation}
\mathcal{F}(\mathcal{K}_4)= \text{min}_{\mathcal{P}_{a_i|A_i}} S(\lbrace \mathcal{P}_{a_i|A_i}\rbrace).
\end{equation}
In order to a lower bound for $\mathcal{F}$ as a function of LGI-violation $\mathcal{K}_4$, we use the operator inequality approach given in Refs. \cite{Kani2,tavakoli'18}. Rewriting $\mathcal{K}_4$ in terms of an operator $W_{ia_i}$ and $\mathcal{P}_{a_i|A_i}$, where $W_{ia_i}= \frac{1}{2} [(-1)^{a_i} (\mathcal{P}_{B_1} + (-1)^{i-1} \mathcal{P}_{B_2})]$, $\mathcal{P}_{B_i}\equiv \mathcal{P}_{0|B_i}-\mathcal{P}_{1|B_i}$, we have,
\begin{equation}
\mathcal{K}_4 =  \sum_{i,a_i} \text{Tr}[W_{ia_i}\mathcal{P}_{a_i|A_i}].
\end{equation}
Let us now define another operator $K_{ia_i}(\mathcal{P}_{B_1}, \mathcal{P}_{B_2}) \equiv \Lambda^{\dagger}(\mathcal{P}_{B_1}, \mathcal{P}_{B_2})[\mathcal{P}_{a_i|A_i}^{\text{ideal}}] $. 
Considering an operator inequality of the form,
\begin{equation}
K_{ia_i}(\mathcal{P}_{B_1}, \mathcal{P}_{B_2}) \geq s W_{ia_i} + \mu_{ia_i} (\mathcal{P}_{B_1}, \mathcal{P}_{B_2}) \mathbb{I}
\end{equation}
with $s$ and $\mu_{ia_i}(\mathcal{P}_{B_1}, \mathcal{P}_{B_2})$ being real coeficients, the average fidelity $S$ (and hence $\mathcal{F}$) can be lower bounded as,
\begin{eqnarray}
\mathcal{F} \geq S &\geq & \frac{1}{4}\sum_{i,a_i} F(\mathcal{P}_{a_i|A_i}^{\text{ideal}}, \Lambda \left[ \mathcal{P}_{a_i|A_i}\right] \nonumber \\
&=& \frac{1}{4}\sum_{i,a_i} Tr[K_{ia_i}\mathcal{P}_{a_i|A_i}] \nonumber \\
&\geq & \frac{s}{4} \sum_{i,a_i} Tr[W_{ia_i}\mathcal{P}_{a_i|A_i}] +\frac{1}{4} \sum_{i,a_i} \mu_{ia_i} \nonumber \\
&=& \frac{s}{4}\mathcal{K}_4 + \mu
\end{eqnarray}
where $\mu \equiv 1/4 ~min_{\mathcal{P}_{B_1}, \mathcal{P}_{B_2}}\sum_{i,a_i} \mu_{ia_i} (\mathcal{P}_{B_1}, \mathcal{P}_{B_2})$.

Since we can use Jordan's lemma to write the observables of Alice in a block-diagonal form with block of size at most $2 \times 2$, it suffices for our purpose to focus on each block as in Ref. \cite{Kani2} to find the constants $s$ and $\mu$ for lower bounding the fidelity. The goal of the robustness analysis is to obtain a self-testing bound as tight as possible. Following Ref. \cite{tavakoli'18},  adoption of the dephasing channel $\Lambda$, as an extraction map  suffices to achieve our aim of finding an optimal self-testing bound.
\begin{equation}
\Lambda_\theta (\rho) = \frac{1+ \xi(\theta)}{2}\rho + \frac{1- \xi(\theta)}{2} \Gamma (\theta) \rho \Gamma(\theta);
\end{equation}
where $\xi(\theta) \in [-1,1] $ and
\begin{align}
\Gamma(\theta)&=\left\{
\begin{array}{lr}
\sigma_z &  \quad  \text{for} \quad  \theta \in \left[0, \frac{\pi}{4}\right]   \\
\sigma_x  &  \quad  \text{for} \quad   \theta \in \left(\frac{\pi}{4}, \frac{\pi}{2}\right] \\
\end{array} \nonumber
\right.
\end{align}

For the interval $0\leq\theta\leq \pi/4$
\begin{eqnarray}
K_{10} &=& \Lambda^\dagger[\mathcal{P}_{0|A_1}] \nonumber \\
&=& \frac{1+ \xi(\theta)}{2}\left[ \frac{\mathbb{I}+\sigma_z}{2}\right] + \frac{1- \xi(\theta)}{2}\left[ \sigma_z \frac{\mathbb{I}+\sigma_z}{2} \sigma_z \right] \nonumber \\
&=& \frac{\mathbb{I}+\sigma_z}{2} ,\nonumber
\end{eqnarray}
and similarly 
$K_{20} =  \frac{\mathbb{I}+\xi(\theta)\sigma_x}{2}$ , $K_{21} =  \frac{\mathbb{I}-\xi(\theta)\sigma_x}{2}$,  $K_{11} =  \frac{\mathbb{I}-\sigma_z}{2}$. \\
whereas, for the interval $\pi/4\leq\theta\leq \pi/2$,\\

~~~~~~~~$K_{10}=\frac{\mathbb{I}+\xi(\theta)\sigma_z}{2}$,~~~~~~~~~~~~ $K_{20}= \frac{\mathbb{I}+\sigma_x}{2}$, \\

~~~~~~~~$K_{11}=\frac{\mathbb{I}-\xi(\theta)\sigma_z}{2}$,~~~~~~~~~~~~ $K_{21}= \frac{\mathbb{I}-\sigma_x}{2}$. \\

We can, without loss of generality, represent the two measurements in the x-z plane as,
\begin{eqnarray}
\mathcal{P}_{B_1} &=& \cos(\theta) \sigma_z + \sin(\theta) \sigma_x \nonumber \\
\mathcal{P}_{B_2} &=& \cos(\theta) \sigma_z - \sin(\theta) \sigma_x 
\end{eqnarray}
and therefore, 
$W_{10} =  \cos(\theta) \sigma_z$, $W_{20} =  \sin(\theta) \sigma_x$, and
$W_{21} = - \sin(\theta) \sigma_x$, $W_{11} = - \cos(\theta) \sigma_z$.

We analyze the operator inequality  in the interval $0\leq\theta\leq \pi/4$ and $\pi/4\leq\theta\leq \pi/2$ separately. One can see that the effective number of  inequalities may be reduced since there are symmetries in the expression of $K_{ia_i}$ and $W_{ia_i}$, and hence, without loss of generality we can choose $\mu_o\equiv \mu_{20}=\mu_{21}$ and $\mu_e\equiv \mu_{10}=\mu_{11}$.
After simplication we get,
\begin{eqnarray}
\mu_e &\leq & 1-s\cos(\theta), \nonumber\\
\mu_e &\leq & s\cos(\theta).
\end{eqnarray}
and,
\begin{eqnarray}
\mu_o &\leq & \frac{1}{2} + \frac{1}{2} \xi(\theta)- s \sin(\theta) ,\nonumber \\
\mu_o &\leq & \frac{1}{2} - \frac{1}{2} \xi(\theta)+ s \sin(\theta).
\end{eqnarray}
In order to obtain the strongest bound, we have to choose the largest value of $\mu_o$ and $\mu_e$ consistant with their respective constraints, {\it i.e.}, in the first interval,
\begin{align}
\mu_e &= \text{min}\left\lbrace 1-s\cos(\theta), s\cos(\theta)\right\rbrace , \nonumber \\
\mu_o &= \text{min}\left\lbrace \frac{1}{2} + \frac{1}{2} \xi(\theta)- s \sin(\theta), \frac{1}{2} - \frac{1}{2} \xi(\theta)+ s \sin(\theta)\right\rbrace .
\end{align}
A similar procedure in the second interval leads to,
\begin{align}
\mu_o &= \text{min}\left\lbrace 1-s\sin(\theta), s\sin(\theta)\right\rbrace , \nonumber \\
\mu_e &= \text{min}\left\lbrace \frac{1}{2} + \frac{1}{2} \xi(\theta)- s \cos(\theta), \frac{1}{2} - \frac{1}{2} \xi(\theta)+ s \cos(\theta)\right\rbrace  .
\end{align}
The expressions in the two intervals are related to each other by the transfomations $\mu_o\leftrightarrow \mu_e$ and $\sin(\theta)\leftrightarrow \cos(\theta)$.

Hence, the lower bound in fidelity becomes,
\begin{equation}
\mathcal{F}(\mathcal{K}_4)\geq \frac{s}{4} \mathcal{K}_4 + \text{min}_{\mathcal{P}_{B_1}, \mathcal{P}_{B_2}} \mu(\mathcal{P}_{B_1}, \mathcal{P}_{B_2}). 
\end{equation}
where $\mu(\mathcal{P}_{B_1}, \mathcal{P}_{B_2})=(\mu_e+\mu_o)/2$. To compute this value, we fix  $s = \frac{1+ \sqrt{2}}{2}$, and choosing the dephasing function as $\xi(\theta) = \text{min}\lbrace 1, 2s \sin(\theta)\rbrace$ in the interval $\theta\in[0,\pi/4]$, and $\xi(\theta) = \text{min}\lbrace 1, 2s \cos(\theta)\rbrace$ in the interval $\theta\in (\pi/4,\pi/2]$. After simplification, we get, $\mu = \frac{2-\sqrt {2}}{4}$ which gives the lower bound,
\begin{equation}\label{robustness1}
\mathcal{F}(\mathcal{K}_4)\geq  \frac{(1+ \sqrt{2})}{8}\mathcal{K}_4 + \frac{2-\sqrt {2}}{4}. 
\end{equation}
This provides robust self-testing of Alice's measurements. Clearly, the maximal quantum violation of the LGI, {\it i.e.}, $\mathcal{K}_4=2\sqrt{2}$ implies $\mathcal{F}(\mathcal{K}_4)=1$, which suggests that the measurements must be the ideal ones.
For $\mathcal{K}_4=2$, $\mathcal{F}(\mathcal{K}_4)\ge 3/4$. This bound can be obtained by $A_1=A_2=B_1=B_2=\sigma_z$. Therefore, we see that our bound is optimal.

We now quantify the average fidelity of the measurements on Bob's side with respect to the ideal ones: $\mathcal{S}^{\prime}(\lbrace \mathcal{P}_{b_i|B_i} \rbrace) = \text{max}_{\Lambda} \sum_{i,b_i} F((\mathcal{P}_{b_i|B_i})^{\text{ideal}}\Lambda [\mathcal{P}_{b_i|B_i}])/4$, where $\Lambda$ must be a unital channel. Now, define,
\begin{equation}
\mathcal{F}^{\prime}(\mathcal{K}_4) = \text{min}_{\lbrace \mathcal{P}_{b_i|B_i} \rbrace} \mathcal{S}^{\prime}(\lbrace \mathcal{P}_{b_i|B_i} \rbrace).
\end{equation}
First, we rewrite $\mathcal{K}_4 =  \sum_{i,b_i} Tr[\mathcal{P}_{b_i|B_i} Z_{ib_i}]$ where  $Z_{ib_i}= \frac{1}{2} (-1)^{b_i}[\mathcal{P}_{A_1} + (-1)^{i-1}\mathcal{P}_{A_2}] $ and $\mathcal{P}_{A_i} \equiv \mathcal{P}_{0|A_i} - \mathcal{P}_{1|A_i}$.

Let us take an operator inequality of the form,
\begin{equation}
K_{ib_i}(\lbrace \mathcal{P}_{A_1}, \mathcal{P}_{A_2} \rbrace) \geq s Z_{ib_i} + \mu_{ib_i} (\lbrace \mathcal{P}_{A_1}, \mathcal{P}_{A_2}\rbrace) \mathbb{I},
\end{equation}
with $K_{ib_i} = \Lambda ^{\dagger}[\mathcal{P}_{b_i|B_i}^{\text{ideal}}]$. Similar to the previous case,
\begin{equation}
\mathcal{F}^{\prime}(\mathcal{K}_4) \geq \text{min}_{\mathcal{P}_{A_1}, \mathcal{P}_{A_2}}\frac{1}{4}\sum_{i,b_i} Tr[K_{ib_i} \mathcal{P}_{b_i|B_i} ].
\end{equation}
Using the same map and the same technique, we  have $\xi(\theta)= \text{min} \lbrace 1, 2s \sin(\theta)\rbrace$  in interval $\theta\in[0,\pi/4]$ and $\xi(\theta) = \text{min}\lbrace 1, 2s \cos(\theta)\rbrace$ $\theta\in (\pi/4,\pi/2]$. 
If we repeat the same procedure as in Alice's measurements, we  get the bound, $\mathcal{F}^{\prime}(\mathcal{K}_4)\geq  \frac{(1+ \sqrt{2})}{8}\mathcal{K}_4 + \frac{2-\sqrt {2}}{4}$. This provides robust self-testing of Bob's measurements which is again optimal.

\begin{figure}
	\begin{center}
		\includegraphics[width=0.40\textwidth]{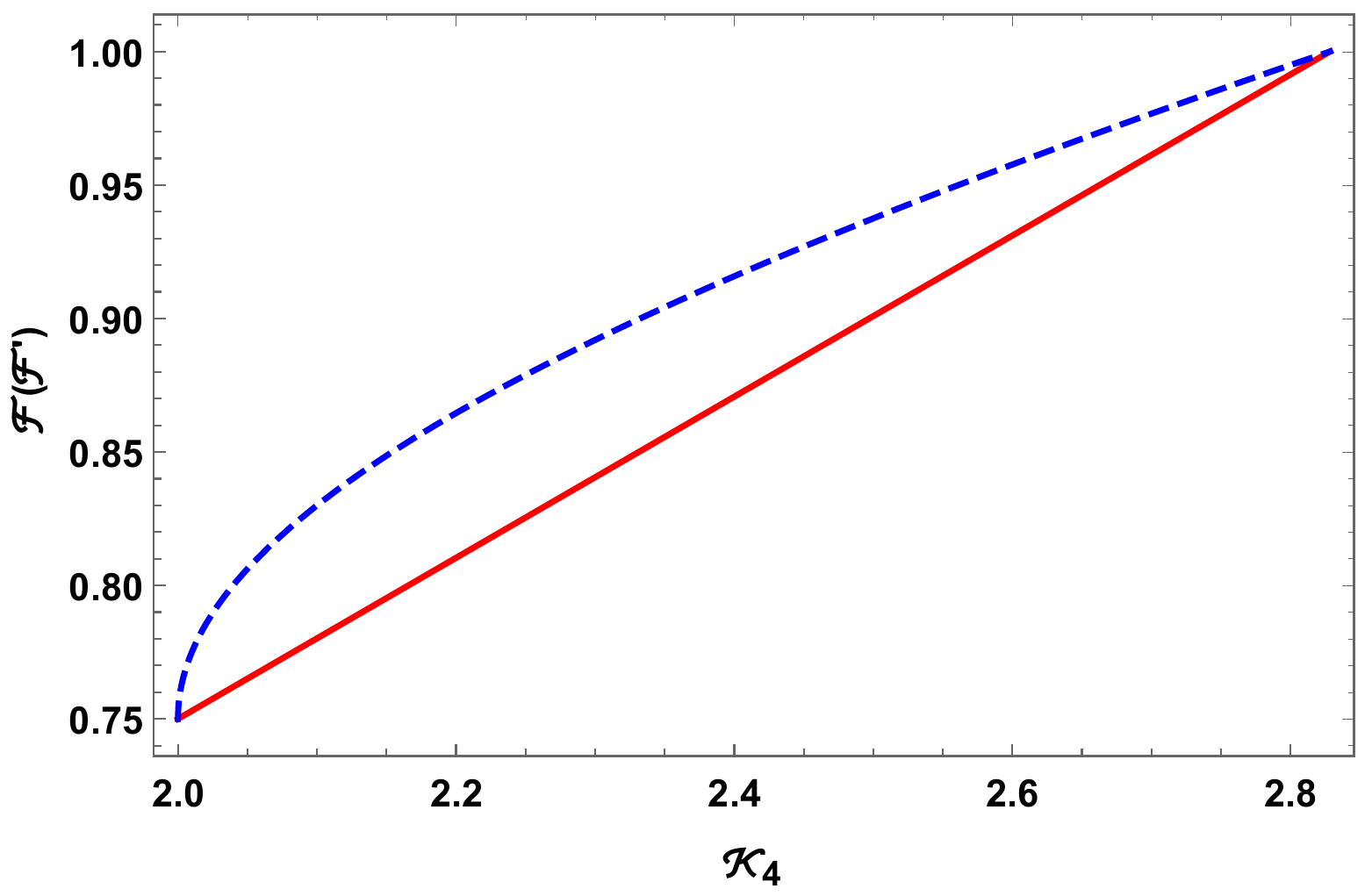}
		\caption{\textcolor{black}{The average fidelity $\mathcal{F}$ ($\mathcal{F^{'}}$), for Alice's (Bob's) measurement  as a function of $\mathcal{K}_4$ is plotted. The red (solid) line represents the lower bound of average fidelity obtained analytically from Eq. \eqref{robustness1}. The blue (dotted) line is the average fidelity under dephasing noise.}\label{Robust}}
	\end{center}
\end{figure}

In order to provide an operational analysis of robustness, let us choose a dephasing channel, $\Lambda_\theta (\rho) = \frac{1+ \xi(\theta)}{2}\rho + \frac{1- \xi(\theta)}{2} \sigma_z \rho \sigma_z$. It keeps the measurement, $A_1 = \sigma_z$ as it is and dephase the measurement, $A_2 =  \sigma_x$ to $\xi(\theta)\sigma_x$. Let us consider Bob's measurements as, $B_1 = \cos(\phi) \sigma_z + \sin(\phi) \sigma_x$ and $B_2 = \cos(\phi) \sigma_z - \sin(\phi) \sigma_x.$
Taking $\xi(\theta)= \tan(\phi)$, we get,
\begin{equation}
\mathcal{K}_4 = 2 \sqrt{1+\tan^2(\phi)} ~~~~~~\text{and} ~~~~~~\mathcal{F} = \frac{1}{4}(3+ \tan(\phi)).\nonumber
\end{equation}
The above expressions provide a parametric curve as a function of $\phi \in [0,\pi/4]$, which is the dashed (blue) line in Fig.-\ref{Robust}. The straight (red) line represents the lower bound of fidelity as given by Eq.\eqref{robustness1}. A straightforward calculation shows that the robustness analysis for Bob's measurements($\mathcal{F}^{\prime}$)  produces a  matching parametric curve.

\section{Conclusions}
With the rapid development of quantum technologies, it is important to first characterize and certify various quantum devices. Among various certification 
protocols, self-testing protocols are designed for the purpose of certifying quantum components from the observed statistics under a set of minimal assumptions. Self-testing is regarded to be more resource-efficient, and requires lesser trust on measurement devices. Previously, self-testing of measurements has been performed mainly employing nonlocal spatial quantum correlations \cite{my'04,Col,supic2018,supic'19,bancal'18,renou'18}, as exhibited by the violation of Bell-type inequalities, which require entanglement, a costly resource. Schemes of self-testing  measurements without entanglement proposed earlier require either dimensional restrictions \cite{tavakoli'18, tavakoli'15}, or other stringent assumptions like compatibility  and projectivity of the measurements \cite{kishor'19,kishor'19(1)} in case of approaches based on contextuality, though it may be possible to relax some of these assumptions for local measurements in three and higher dimensions \cite{saha'20}.  

In this work our purpose is to self-test binary measurements with dichotomic settings and outcomes without using entangled states. To this end we have exploited another fundamental property of quantum mechanics, {\it viz.}, temporal quantum correlations exhibited by violation of Leggett-Garg inequalities together with violation of predictability \cite{LG'85, emary'14, kofler'13, brukner'04, rand'16}. We have presented a  scheme of self-testing of binary Pauli measurements through LGI violation using a minimal assumption that is inevitable in the present context. We have shown how the maximum quantum violation of the four-term LGI along with the NSIT condition can be used to provide a self-testing statement of  binary measurements without entanglement, which requires neither compatibility or projectivity, nor any dimensional restriction. Moreover, we have formulated the robustness bound of our self-testing protocol, and analysed it with an example of dephasing noise.

Before concluding, we would like to highlight certain salient features of our analysis. First, it is evident from our analysis that the proposed self-testing protocol for binary qubit measurements does not depend on the input state. Secondly, though in the present work we have not discussed self-testing of quantum states, once the measurement devices are characterized through our approach of self-testing  through the  LGI framework, any subsequent certification of quantum states using such devices may reduce essentially to the task of decoherence control. Finally, the NSIT condition used here is experimentally implementable, as shown recently, for a variety of states \cite{expnsit}. Experimental viability of our protocol is ensured in the backdrop of several recent LG tests \cite{knee'12,robens'15,knee'16,expnsit, ku'19,shayan'19,spee'20}.

\begin{acknowledgments}
AGM would like to thank Debarshi Das and Debashis Saha for helpful comments. SM acknowledges the Ministry of Science and Technology in Taiwan (Grant no. 110-2811-M-006 -501). CJ thanks Huan-Yu Ku for discussions and acknowledges the financial support from the Ministry of Science and Technology, Taiwan (Grant No.  MOST 108-2811-M-006-516) and the Foundation for Polish Science through the First Team project (No First TEAM/2017-4/31). ASM acknowledges support from the project DST/ICPS/QuEST/Q98 from the Department of Science and Technology, India.
\end{acknowledgments}

\appendix
\section{Maximal violation of the LGI with POVMs.}\label{Appa}
Here we obtain the maximal of the Leggett-Garg inequality for POVMs \cite{busch'91,das'18} which are a set of positive operators that add to identity, {\it i. e.},  $E \equiv \{ E_i | \sum_i E_i = \mathbb{I}, 0 <E_i \leq \mathbb{I} \}$. Each of the operators $E_i$, called effect operator determines the probability  $\text{Tr}[\rho E_{i}]$ of obtaining the $i^{\text{th}}$ outcome when applied on the state $\rho$. The most general POVM with two outcomes are defined by two parameters,-- sharpness parameter and biasedness parameter. Let us  consider that $\gamma_{a}, \gamma_{b}$ are the biasedness parameter and $\lambda_i$ and $\mu_j$ are the sharpness parameters for Alice's $i$-th and Bob's $j$-th measurements, respectively. The most general effect operators with two outcomes can be written as, 
\begin{eqnarray}
&&E^{\lambda_i}_{a_i} = \lambda_i \mathcal{P}_{a_i|A_i} + (1\pm \gamma_{a} -\lambda_i) \frac{\mathbb{I}_2}{2}, \nonumber\\
&&E^{\mu_j}_{b_j} = \mu_j \mathcal{P}_{b_j|B_j} + (1 \pm \gamma_{b}-\mu_j) \frac{\mathbb{I}_2}{2}
\end{eqnarray}
The post measurement state can be derived using generalized von Neumann-L\"{u}ders transformation rule, $\dfrac{\sqrt{E^{\lambda_i}_{a_i}} \rho \sqrt{E^{\lambda_i}_{a_i}}}{\text{Tr}[E^{\lambda_i}_{a_i} \rho]}$.
\\
Now, in order to derive the LGI we shall follow the same scenario and procedure as discussed in the main text. The two-time joint probability is obtained using Baye's rule,
\begin{eqnarray}
&&P(a_i,b_j \mid A_i,B_j) = P(a_i \mid A_i) P(b_j\mid a_i, A_i,B_j ) \nonumber \\
&=& Tr\left[ E^{\lambda_i}_{a_i} \rho_{in}\right] Tr\left[ E^{\mu_j}_{b_j} \dfrac{\sqrt{E^{\lambda_i}_{a_i}} \rho \sqrt{E^{\lambda_i}_{a_i}}}{\text{Tr}[E^{\lambda_i}_{a_i} \rho]}\right] .
\end{eqnarray}
The two-time correlation is defined as,
\begin{eqnarray}
\mathcal{C}_{ij} =\sum_{a_i,b_j} (-1)^{a_i\oplus b_j} P(a_i,b_j \mid A_i,B_j),
\end{eqnarray}
where $\oplus$ denotes addition modulo $2$. 
The four-term LGI in terms of the above correlators is given by,
\begin{equation}
\mathcal{K}_4 = C_{11} + C_{21} + C_{22} - C_{12} \leq 2
\end{equation}
For a general input state $\rho_{in}=\frac{\mathbb{I}+\hat{n}.\vec{\sigma}}{2}$, the first term  (probability obtained by Alice) reduces to, 
\begin{equation}
Tr\left[ E^{\lambda_i}_{a_i} \rho_{in}\right] = \frac{1}{2}((1\pm \gamma_a) + (-1)^{a_i}\lambda_i\hat{a_i}.\hat{n}) \nonumber
\end{equation}
Similarly, one can calcultate $Tr\left[ E^{\mu_j}_{b_j} \dfrac{\sqrt{E^{\lambda_i}_{a_i}} \rho \sqrt{E^{\lambda_i}_{a_i}}}{\text{Tr}[E^{\lambda_i}_{a_i} \rho]}\right]$
One can now calculate $\mathcal{K}_{4}$ and maximize it numerically. It can be found that the maximum quantum value of $\mathcal{K}_{4}$ is $2\sqrt{2}$ when all $\vert \hat{a_i}.\hat{b_j}\vert=\frac{1}{\sqrt{2}}$ along with $\lambda_i=\mu_j=1$ and $\gamma_{a}=\gamma_{b}=0$. Hence, the  maximum value of LGI can only be achieved if Alice and Bob perform sharp-projective measurements.

\section{Details of proof of Theorem 1.}\label{Appb}
Appending an ancilla qubit prepared in the state $\left|0 \right \rangle$, let us show that there exists an isometry $\Phi$ defined by the map,
\begin{align}
\Phi \left|2m,0 \right \rangle &\rightarrow  \left| 2m, 0   \right \rangle  \nonumber  \\
\Phi \left|2m+1,0 \right \rangle &\rightarrow  \left| 2m, 1  \right \rangle
\end{align} 
such that 
\begin{eqnarray}
&&\Phi\left( B_j \frac{\mathcal{P}_{a_i|A_i}\rho_{in}\mathcal{P}_{a_i|A_i}^{\dagger}}{\text{Tr}\left[ \mathcal{P}_{a_i|A_i}\rho_{in}\mathcal{P}_{a_i|A_i}^{\dagger}\right]} \otimes \left|0 \right \rangle \left \langle 0 \right | \right) \Phi^{\dagger} \nonumber \\
&=&B^{\texttt{ideal}}_j \left|\psi^{\texttt{ideal}}_{a|A_i}\right \rangle \left \langle\psi^{\texttt{ideal}}_{a|A_i} \right |\otimes \left| \texttt{junk} \right \rangle \left \langle \texttt{junk}\right |\nonumber
\end{eqnarray}
holds.
\\
We present here the calculation for one term (say, $\Phi\left( B_1 \frac{\mathcal{P}_{a_1|A_1}\rho_{in}\mathcal{P}_{a_1|A_1}^{\dagger}}{\text{Tr}\left[ \mathcal{P}_{a_1|A_1}\rho_{in}\mathcal{P}_{a_1|A_1}^{\dagger}\right]} \otimes \left|0 \right \rangle \left \langle 0 \right | \right) \Phi^{\dagger}$) explicitly. Other terms can be calculated in a similar fashion. 
\\
Choosing the eigenbasis of $A_1$ as the computational basis, from lemma $1$ and $2$, it follows that the post-measurement states of Alice can be written as  
$\frac{\mathcal{P}_{a_i|A_i}\rho_{in}\mathcal{P}_{a_i|A_i}^{\dagger}}{\text{Tr}\left[ \mathcal{P}_{a_i|A_i}\rho_{in}\mathcal{P}_{a_i|A_i}^{\dagger}\right]} = \oplus_m p_m \left|\psi^m_{a|A_1}\right \rangle \left \langle \psi^m_{a|A_1}\right | $, with $\left|\psi^m_{0|A_1}\right \rangle= \left| 2m \right \rangle $
and $\left|\psi^m_{1|A_1}\right \rangle= \left| 2m +1 \right \rangle $.
The measurement $B_1$ on Bob's side is $ \oplus_n (\sigma_x^n+ \sigma_z^n)/\sqrt{2} $.
Expanding the sum  for $a=0$ leads to
\begin{widetext}
	\begin{align}
	&     \oplus_n \frac{\sigma_x^n+ \sigma_z^n}{\sqrt{2}} \oplus_m p_m  \left| 2m \right \rangle \left \langle 2m \right | \nonumber \\
	&= \bigg( \frac{\ketbra{0}{0}-\ketbra{1}{1}+\ketbra{0}{1}+\ketbra{1}{0}}{\sqrt{2}}+\frac{\ketbra{2}{2}-\ketbra{3}{3}+\ketbra{2}{3}+\ketbra{3}{2}}{\sqrt{2}}+\cdots 
	\nonumber \\ 
	&+\frac{\ketbra{2m}{2m}-\ketbra{2m+1}{2m+1}+\ketbra{2m}{2m+1}+\ketbra{2m+1}{2m}}{\sqrt{2}} \bigg)  \cdot \left( p_0 \ket{0}\bra{0} + p_1 \ket{2}\bra{2} + \cdots +  p_m \ket{2m}\bra{2m} \right)  \nonumber \\
	&=\bigg( p_0 \frac{\ket{0}+\ket{1}}{\sqrt{2}}\bra{0}+ p_1 \frac{\ket{2}+\ket{3}}{\sqrt{2}}\bra{2}+\cdots +p_m \frac{\ket{2m}+\ket{2m+1}}{\sqrt{2}}\bra{2m} \bigg) 
	\end{align}	
	Now, for the map defined above,
	\begin{align}
	&\Phi \left(\oplus_n \frac{\sigma_x^n+ \sigma_z^n}{\sqrt{2}} \oplus_m p_m  \left| 2m \right \rangle \left \langle 2m\right | \otimes \ket{0}\bra{0} \right) \Phi^{\dagger}\nonumber \\
	&=\frac{\ket{0}+\ket{1}}{\sqrt{2}}\bra{0} \otimes  \oplus_m p_m  \left| 2m \right \rangle\left \langle 2m\right | \nonumber \\
	&=B^{\texttt{ideal}}_1 \left|\psi^{\texttt{ideal}}_{0|A_1}\right \rangle \left \langle\psi^{\texttt{ideal}}_{0|A_1} \right |\otimes \left| \texttt{junk} \right \rangle \left \langle \texttt{junk}\right |
	\end{align}
	with $\left| \texttt{junk} \right \rangle \left \langle \texttt{junk}\right |= \sum_m p_m  \left| 2m \right \rangle \left \langle 2m\right |$.
	\\
	Next, for $a=1$, we have
	\begin{align}
	&     \oplus_n \frac{\sigma_x^n+ \sigma_z^n}{\sqrt{2}} \oplus_m p_m  \left| 2m +1 \right \rangle \left \langle 2m+1 \right | \nonumber \\
	&=\bigg( p_0 \frac{\ket{0}-\ket{1}}{\sqrt{2}}\bra{0}+ p_1 \frac{\ket{2}-\ket{3}}{\sqrt{2}}\bra{2}+\cdots +p_m \frac{\ket{2m}-\ket{2m+1}}{\sqrt{2}}\bra{2m} \bigg) 
	\end{align}
	Hence,  for the map defined above, 
	\begin{align}
	&\Phi \left(\oplus_n \frac{\sigma_x^n+ \sigma_z^n}{\sqrt{2}} \oplus_m p_m  \left| 2m+1 \right \rangle\left \langle 2m+1 \right | \otimes \ket{0}\bra{0} \right) \Phi^{\dagger}\nonumber \\
	&=\frac{\ket{0}-\ket{1}}{\sqrt{2}}\bra{0} \otimes  \oplus_m p_m  \left| 2m \right \rangle\left \langle 2m \right | \nonumber \\
	&=B^{\texttt{ideal}}_1 \left|\psi^{\texttt{ideal}}_{1|A_1}\right \rangle  \left \langle\psi^{\texttt{ideal}}_{1|A_1} \right |\otimes \left| \texttt{junk} \right \rangle \left \langle \texttt{junk}\right | .
	\end{align}
\end{widetext}
\end{document}